# Disorder-induced significant enhancement in magnetization of ball-milled $Fe_2CrGa$ alloy

H. G. Zhang,[1] C. Z. Zhang,[2] W. Zhu,[1] E. K. Liu,[1] W. H. Wang,[1] H. W. Zhang,[1] J. L. Cheng,[1] H. Z. Luo,[3] G. H. Wu[1, a)]

*1 Beijing National Laboratory for Condensed Matter Physics, Institute of Physics, Chinese Academy of Sciences, Beijing 100190, People's Republic of China*

*2 School of Physical Science and Technology, Inner Mongolia University, Hohhot 010021, People's Republic of China*

*3 School of Material Science and Engineering, Hebei University of Technology, Tianjin 300130, People's Republic of China*

[a)] Author to whom correspondence should be addressed. Electronic mail: ghwu@aphy.iphy.ac.cn.

**Abstract:** A new disordered atom configuration in $Fe_2CrGa$ alloy has been created by ball-milling method. This leads to a significant enhancement of the magnetic moment up to 3.2~3.9 $\mu_B$ and an increase of Curie temperature by about 200 K, compared with the arc-melt samples. Combination of first-principles calculations and experimental results reveals that $Fe_2CrGa$ alloy should crystallize in $Hg_2CuTi$ based structure with different atomic disorders for the samples prepared by different methods. It is addressed that magnetic interactions play a crucial role for the system to adopt such an atomic configuration which disobeys the empirical rule.

Heusler alloys with formula $X_2YZ$ have been intensively studied for decades due to their abundant physical properties and great potential in applications.[1-7] Recently, theoretical investigations predicted that, like some Co-based Heusler alloys,[3] Fe-based Heusler alloys such as $Fe_2CrZ$ (Z = Ge, Sn and Sb) also have the possibility to be half-metallic ferromagnets,[8,9] if the alloys crystallized in $L2_1$ structure. The half-metallic ferromagnets have great potential of usage in the field of spintronics.[10] However, the atomic configurations and magnetic structures in some $Fe_2Cr$-based Heusler alloys are still questionable currently. In Heusler alloys, the atomic configuration obeys such an empirical rule:[11-13] the metallic atoms with more valence electrons occupy the nearest neighbour sites to the main group elements preferentially. According to this rule, $Fe_2CrGa$ alloy should adopt a $L2_1$ structure, though the theoretical calculation based on $L2_1$ gives a rather different magnetic moment from the experimental results.[9,14] Zhu W. *et al*.[15] assumed that the actual structure of $Fe_2CrGa$ should be $Hg_2CuTi$ rather than $L2_1$ type, and their calculated magnetic moment based on $Hg_2CuTi$ is very close to the experimental result. Umetsu R. Y. *et al*.[16] found that the structural and magnetic properties of $Fe_2CrGa$ are very sensitive to heat treatment. Besides, a precipitation behavior has also been found in these works. Due to these uncertainties, a systematic investigation on the atom configuration and the corresponding magnetic structure for this alloy is imperative.

In this letter, we report the very high magnetic moments and Curie temperatures obtained in the ball-milled $Fe_2CrGa$ samples. The system is found to crystalline in $Hg_2CuTi$ based structure with partial disordering. Magnetic interactions are thought to play a determinative role in this system for its violation of the experiential rule.

$Fe_2CrGa$ ingots were prepared by arc-melting in an argon atmosphere. Then they were milled

in a planetary ball mill for 21 hours. The as-milled powder was compacted into pellets and annealed at different temperatures before quenched into ice water. X-ray diffraction (XRD) and selected area electron diffraction (SAED) were performed for the structure of the samples. Magnetization measurements were carried out on vibrating sample magnetometer (VSM) and superconducting quantum interference device (SQUID). Theoretic calculations of the magnetic moment and total energy were performed based on the first-principles KKR-CPA-LDA method.[15,17-19]

Figure 1 shows the XRD patterns and the annealing temperature dependence of the lattice constant. All the samples state in a pure body centered cubic (bcc) phase and no sign of second phase can be observed. The atomic order could not be determined by the indistinct superlattice reflection of the XRD patterns, duing to the small difference between the atomic scattering factors.[16] However, it is reasonable to believe that a serious atomic disorder and possible clusters might occur during the ball-mill process.[20] The lattice constant of the as-milled sample is relatively large (5.83 Å) and drastically decreases up to 673 K, then stabilized around 5.81 Å in the range of 773 K to 1273 K. It suggests an atom ordering process and elimination of possible clusters, since there is an order-disorder transition occurred at about 770 K according to previous works.[15,16]

Temperature dependences of magnetization (MT) have been measured to indirectly identify the phase quality of these samples.[15,16] As shown in the inset of Fig. 2 (a), the MT curves of the as-milled and 473 K annealed samples clearly show two Curie temperatures ($T_C$), indicating the existence of a second magnetic phase which is similar with the one reported in several works.[21-25] Fig. 2(a) shows the annealing temperature dependence of the matrix $T_C$ with a peak behavior, while the cluster $T_C$ (at higher temperature) is almost unchanged. It suggests the melting of clusters and variation of matrix composition in the annealing temperature range of 300~673 K. After that, the system undergoes an ordering process around 773 K and shows a single $T_C$ in the whole temperature range (5 K~800 K), as shown in the inset of Fig. 2(a). These results indicate that the magnetic clusters can be eliminated effectively by annealing at 773 K or higher temperatures. Therefore, the samples in this work were annealed in two schemes: directly annealed at certain temperatures, or pretreated at 1073 K for 3 days before further annealing processes (denoted as A- and B-samples, respectively).

Fig. 2(b) shows that B-samples present a relatively even $T_C$ behavior, except for the one annealed at 573 K, which has a double-$T_C$ MT curve (inset of Fig. 2(b)). This reveals the specific temperature at which a precipitation occurs in $Fe_2CrGa$. The inset of Fig. 2(b) also shows the MT curve of the sample annealed at 1473 K, which has a single $T_C$ at about 620 K, suggesting an abnormal ordering state in this sample. It is noteworthy that $T_C$ of these ball-milled samples is about 200 K higher than that of arc-melt samples.[15,16] These unusual thermomagnetic behaviors strongly suggest that the ball-milling process has created a different atomic order from the melting method in this system.

Fig. 3 shows the SAED patterns for the 1073 K annealed A-sample and the 573 K annealed B-sample. Though the (200) spots are indistinct, Fig. 3(a) still reveals a pure, partially ordered, B2-like phase in the 1073 K annealed sample.[26,27] However, the B-sample annealed at 573 K shows additional streaks due to the precipitate in this sample, as shown in Fig.3 (b). This confirms the actual phase-separation temperature as discussed above. The high atom order ($L2_1$) of this precipitated sample suggests that the high order states are energy unfavorable in this system and only deviating from the stoichiometric composition could realize it.

The annealing temperature dependences of the saturation magnetic moments ($M_S$) of A- and B-samples are shown in Fig. 4, where the inset shows the MH curves measured at 5 K. The as-milled sample possesses a relatively large $M_S$ (3.6 $\mu_B$), which decreases dramatically to a stable value of about 3.2 $\mu_B$ at 773 K, and then keeps consistent with the values of B-samples. This could prove that the magnetic clusters were eliminated during the atomic ordering process, as shown above in Fig. 1 and Fig. 2. The quite large $M_S$ in the samples annealed at temperatures below 773 K make it reasonable to believe that the clusters consist of Fe-rich composition. On the other hand, the $M_S$ of B-samples are nearly independence with the annealing temperature. There is an exception at 573 K, showing an abnormal large $M_S$, which corresponds to the precipitation occurred at this temperature.

The kink points at 773 K on both curves, showing a minimum of $M_S$, reflect the atom ordering in our ball-milled samples. Since no significant change of the molecular moment at this point, it suggests that the atom ordering at 773 K does not change the nearest neighboring relationship between A/B and C/D magnetic sublattices which, as the discussions below, primarily dominate the molecular moment in $Fe_2CrGa$ alloy. The significant enhancement of $M_S$ by ball milling, in sharp contrast with arc melting, demonstrates that the milling process has induced a different atom configuration and thus a different magnetic structure in this system. The inset of Fig.4 also shows that the B-sample annealed at 1473 K possesses a surprisingly high $M_S$ of about 3.9 $\mu_B$, which will be interpreted by the following theoretical calculations.

So far, various molecular moments have been observed in $Fe_2CrGa$ alloys both experimentally and theoretically.[9,14,16] To solve the divergence, all possible ordering states, as shown in Table 1, have been considered and calculated using CPA method. The magnetic moment of $L2_1$ structure mainly relies on Cr atoms on B sites ($Cr_B$) which forms a ferromagnetic structure with $Fe_A$ and $Fe_C$ sublattices. However, if $Hg_2CuTi$ structure is considered, $Fe_A$ and $Fe_C$ will possess moments of about 1.74 and 2.52 $\mu_B$, respectively, and become the main magnetic contributors. It is also clear to see that all the $Hg_2CuTi$ based B2 structures show larger molecular moments than those of $L2_1$ and the $B2^1_{BD}$ states.

Figure 5 shows the disorder degree dependences of the calculated total energy in $Fe_2CrGa$ system, considering magnetic interactions or not. The calculations were carried out along different disordering paths between $L2_1/Hg_2CuTi$ and six kinds of B2-type configurations. Without magnetism, $L2_1$ structure states in a lower energy level than $Hg_2CuTi$. However, the energy decrement of $Hg_2CuTi$ is much larger than that of $L2_1$ structure if the magnetic interactions are considered. Consequently, the $Hg_2CuTi$ structure becomes more stable than $L2_1$ in an actual magnetic system. These calculations prove that magnetic interactions result in the violation of the empirical rule mentioned above, leading the Fe2CrGa system to crystallize in an Hg2CuTi structure.

Fig. 5(b) indicates that disorder based on $Hg_2CuTi$ structure will result in two even more stable states, $B2_{AD}$ and $B2_{CD}$, which should be considered as the most possible configurations for the $Fe_2CrGa$ alloy. The magnetic moments around these two states are also present in Fig. 5(b). The minimum near $B2_{AD}$ state corresponds to a moment consisted with the arc-melt samples while the other one near $B2_{CD}$ has a moment agreed with our ball-milled samples. These results suggest that the Fe$_2$CrGa alloy should not only crystallize in Hg$_2$CuTi based structures, but also has partial Fe$_A$/Ga$_D$ or Cr$_C$/Ga$_D$ disorders, depending on the preparation methods i.e. arc melting and ball milling.

Additionally, since the arc-melt sample could reach A2 state when annealed at 873 K,[16] it is reasonable to infer that ball-milled sample could also states in A2 structure when annealed at high temperature like 1473 K. Therefore, a series of disordered states from B2$_{CD}$ to A2 had also been calculated, as shown in Fig. 5(b). There is a metastable state near the A2 state, which is absence in the non-magnetic situation. This state corresponds to a magnetic moment of 3.9 $\mu_B$ which is consistent with the value of the sample annealed at 1473 K, indicating that the large moment in this sample comes from the most disorder state induced by ball milling and annealing.

To illustrate the variation of magnetic structure, the atomic moments in Fe$_2$CrGa alloy along the disordering path from Hg$_2$CuTi to A2 structure has been shown in Fig. 6. Based on the discussions above on Table 1, Fig. 6(a) shows that Fe$_A$ and Fe$_B$ are the ferromagnetic contributors in Hg$_2$CuTi structure. While Cr$_C$/Ga$_D$ disordering happens, Cr atoms immigrate to D sites and their moments become antiparallel with Cr$_B$ due to the strong antiferromagnetic exchange interactions. Consequently, as shown in Fig. 6(b), moments on Cr$_D$ are parallel aligned with those of FeA/FeA atoms and provide a positive contribution to the ferromagnetism, inducing the enhancement of magnetization in this system. Fig. 6 shows that Fe atoms maintain this large ferromagnetic contribution of about 2 $\mu_B$ up to A2 state, which results in the large moment in 1473 K annealed sample.

In conclusion, pure ball-milled Fe$_2$CrGa alloy has been obtained by eliminating the Fe rich clusters above 773K. The precipitation in this system has been confirmed to happen at 573K. It has been found that the ball-milled sample has a quite large molecular moment of 3.2~3.9$\mu_B$, which is a significant enhancement compared with arc-melt samples. Highly consistent with the experimental observations, CPA calculations reveal that the Fe$_2$CrGa should crystallize in an Hg$_2$CuTi based structure with partial disordering. Different preparation methods lead the system to take different disordering paths, while magnetic interactions make the system to choose an atomic configuration disobeyed the empirical rule.

This work is supported by the National Natural Science Foundation of China in Grant No. 51031004, 51171206 and 11174352 and National Basic Research Program of China (973 Program, 2010CB619405).

Table Ⅰ. The total energies and magnetic moments calculated by CPA for some particular atomic configurations: $L2_1$, $Hg_2CuTi$, A2 (100% disordered) and B2 (50% disordered). $B2^1_{BD}$ stands for Cr/Ga disorder on B/D sites based on the $L2_1$ structure while $B2^2_{BD}$ represents the Fe/Ga disorder on B/D sites based on the $Hg_2CuTi$ structure. Due to the equivalence of the other B2 states between $L2_1$ and $Hg_2CuTi$, the states of $B2_{AD}$, $B2_{AC}$, $B2_{CD}$ and $B2_{BC}$ are all derived from the $Hg_2CuTi$.

| Structure | Degree of disorder (%) | Total energy (Ry) | Magnetic moment ($\mu_B$) | | |
|---|---|---|---|---|---|
| $L2_1$ | 0 | -11083.75614 | 0.9747 | $Fe_A$ | -0.33 |
| | | | | $Cr_B$ | 1.60 |
| | | | | $Fe_C$ | -0.33 |
| $Hg_2CuTi$ | 0 | -11083.75871 | 2.3504 | $Fe_A$ | 1.74 |
| | | | | $Fe_B$ | 2.52 |
| | | | | $Cr_C$ | -1.69 |
| A2 | 100 | -11083.7526 | 3.93 | | |
| $B2^1_{BD}$ | 50 ($Cr_B/Ga_D$) | -11083.75795 | 0.73 | | |
| $B2_{AD}$ | 50 ($Fe_A/Ga_D$) | -11083.75977 | 2.61 | | |
| $B2_{AC}$ | 50 ($Fe_A/Cr_C$) | -11083.74528 | 2.68 | | |
| $B2_{CD}$ | 50 ($Cr_C/Ga_D$) | -11083.76096 | 3.59 | | |
| $B2^2_{BD}$ | 50 ($Fe_B/Ga_D$) | -11083.75981 | 1.95 | | |
| $B2_{BC}$ | 50 ($Fe_B/Cr_C$) | -11083.74673 | 3.06 | | |

Figures

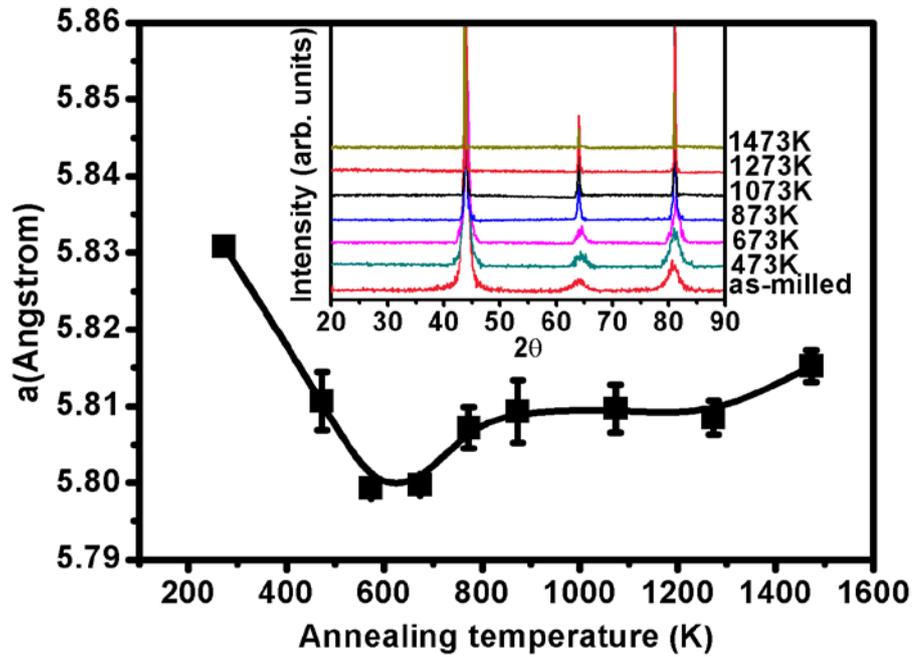

Figure 1. Annealing temperature dependence of lattice constant for ball-milled $Fe_2CrGa$ samples. The inset shows the XRD patterns.

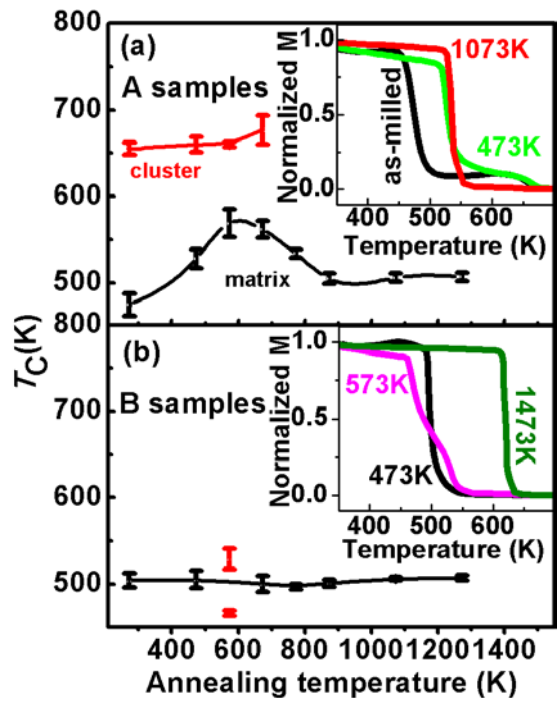

Figure 2. The annealing temperature dependences of Curie temperature of (a) A- and (b) B-samples. The insets show the MT curves.

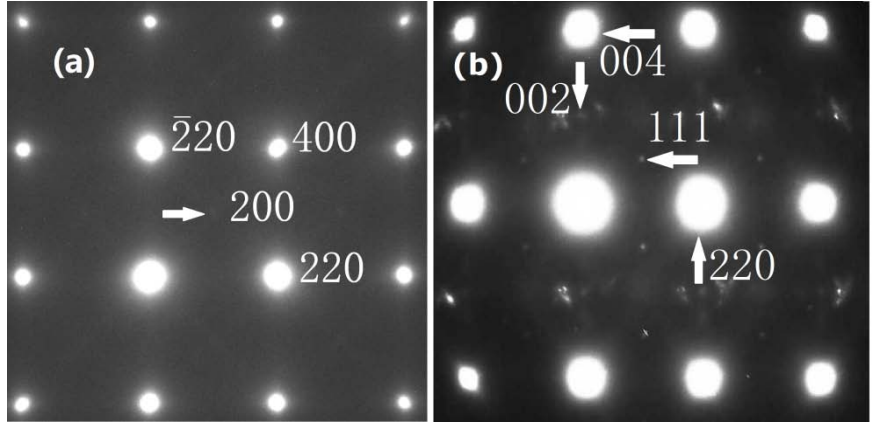

Figure 3. SAED patterns of the (a)1073 K annealed A-sample and (b) 573 K annealed B-sample with the EB parallel to [100] and [110] directions, respectively.

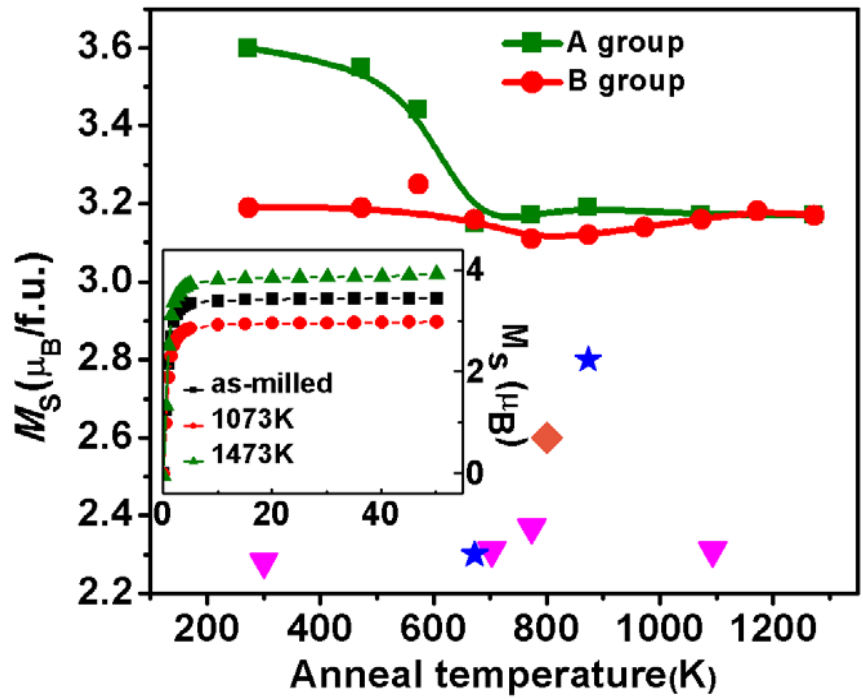

Figure 4. $M_S$ as a function of annealing temperature for A- and B-samples. The inset shows the MH curves measured at 5 K for several samples. For comparison, the molecule moment obtained from the previous works are also shown in the figure.[14-16]

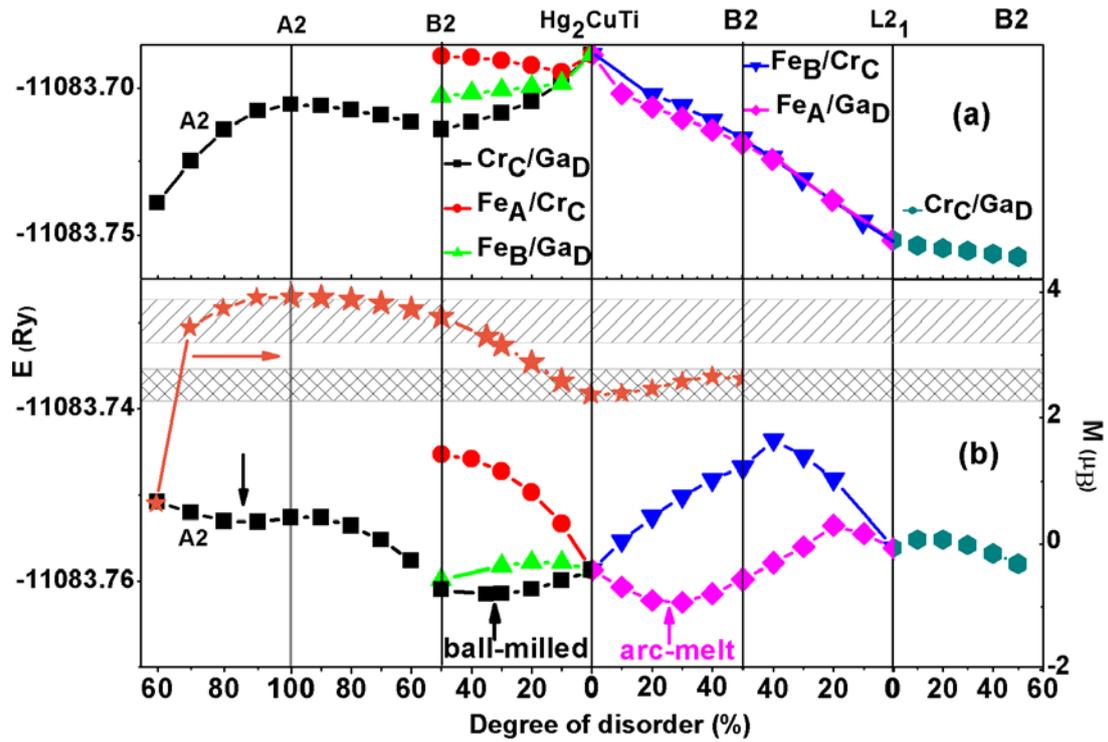

Figure 5. The total energies and magnetic moments (stars) calculated as a function of disordering degree in $Fe_2CrGa$ by CPA, (a) considering without and (b) with the magnetic interactions, respectively. The shadow areas stand for the range of measured magnetic moment in this work (upper one) and in the previous works (lower one).

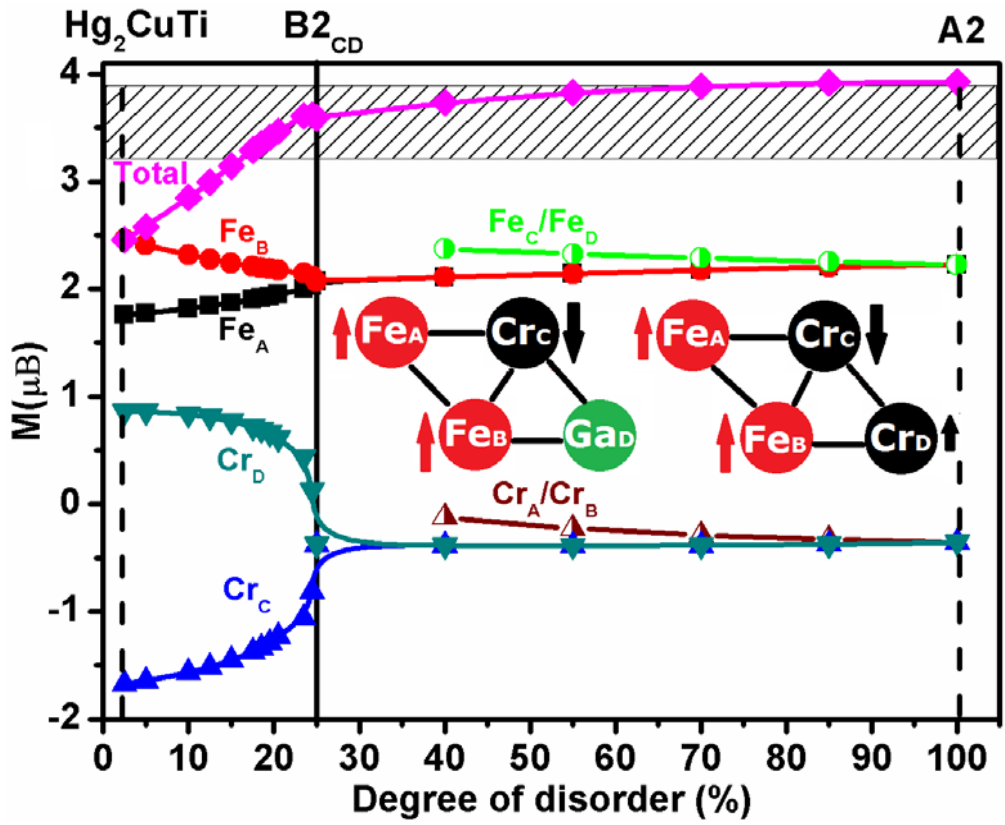

Figure 6. The atomic moments calculated along the disordering path from $Hg_2CuTi$ to A2 structure. The insets show the magnetic structure of $Hg_2CuTi$ (a) and $B2_{CD}$ (b) structure in [011] plane of $Fe_2CrGa$. The shadow area represents the range of measured magnetic moments in this work.